\documentclass{aa}
\usepackage{graphicx}
\begin{document}

   \title{Vacuum nonlinear electrodynamic effects \\
in hard emission of pulsars and magnetars}

   \author{Victor I. Denisov
          \and
Sergey I. Svertilov
          }


   \institute{Physics Department, Moscow State University,\\
    119899, Moscow, Russia
\\
              \email{denisov@srd.sinp.msu.ru ; sis@coronas.ru}
             }

   \date{ {\bf{The shortened version was published
 in A$\&$A 399, L39 (2003)} } }

\titlerunning{Vacuum nonlinear electrodynamic effects}
\authorrunning{V. I. Denisov, S.I. Svertilov}

   \abstract{
The possibilities of observing some nonlinear electrodynamic effects, 
which can be manifested in hard emission of X-ray, gamma ray pulsars 
and magnetars by X-ray and gamma ray astronomy methods are discussed. 
The angular resolution and sensitivity of modern space observatories 
give the opportunity to study the nonlinear electrodynamic effects, 
which can occur in very strong magnetic fields of pulsars 
($B\sim 10^{12}$ G) and magnetars ($B\sim 10^{15}$ G). Such magnetic
field magnitudes are comparable with the typical value of magnetic 
field induction necessary for manifestation of electrodynamics 
non-linearity in vacuum. Thus, near a magneticneutron star the 
electromagnetic emission should undergo nonlinear electrodynamic 
effects in strong magnetic fields (such as bending of rays, fluxes    
dispersing, changing of spectra and polarization states). 
Manifestations of these effects in detected hard emission from magnetic
neutron stars are discussed on the base of nonlinear generalizations     
of the Maxwell equation in vacuum. The dispersion equations for 
electromagnetic waves propagating  in the magnetic dipole field were    
obtained in the framework of these theories.The possibility of 
observing the bending of a ray and gamma ray flux dispersing in the     
neutron star magnetic field are analyzed. The only nonlinear 
electrodynamicseffect, which can be measured principally, is the
effect of gamma ray flux dispersion by the neutron star magnetic           
field. Studying this effect we can also obtain information on the 
nonlinear electrodynamics bending of a ray in the source. The main     
qualitative difference in predictions of different nonlinear 
electrodynamics theories are discussed.
             }
   \keywords{binaries: close --- polarization --- pulsars: general ---
scattering --- stars: neutron --- waves  }

   \maketitle
%

\section{Introduction}

The advanced X-ray and gamma-ray space observatories, such
as the well known Chandra (Van Spreybroeck et al. 1997; CXC 2000
\footnote{Chandra MSFC, and Chandra IPI Teams 2000,
Chandra Proposers Observatory Guide (POG) Electronic Version 2.0 is
available at: http://asc.harvard.edu/udocs/docs/docs.html}),
XMM-Newton (HEASARC 2001
\footnote{High Energy Astrophysics Science Archive
Research Center 2001, "A Portable Mission Count Rate
Simulator (W3PIMMS)", Version 3.1c is available at:
http://heasarc.gsfc.nasa.gov./TOOLS/w3pimms.html}),
and INTEGRAL (Gehrels et al. 1997; Winkler 1999)
permit to observe astrophysical
sources in their hard emission
at the sensitivity levels $\sim 10^{-5}$ phot
sm$^{-2}$s$^{-1}$\ keV$^{-1}$ (for 0.1-10 keV photons) and $\sim
10^{-6}$ phot cm$^{-2}$s$^{-1}$\ keV$^{-1}$ (for 0.1-1.0 MeV photons) for
real time exposures ($\sim 10^6$ s)
with the angular resolution of $\sim
0.1$ arc sec (soft X-rays) and about ten arc min (gamma-rays).
The space missions, which are in
formulation now, such as Constellation X, XEUS, NeXT will improve the
sensitivity by 15-100 times and MAXIM Pathfinder gives a 1000 times
increase in the angular resolution (Weaver, White, $\&$  Tananbaum
2000; White $\&$ Tananbaum 2000; Parmar et al. 1999; Cash, White,
$\&$ Joy 2000).  This gives us an opportunity to study the physical
conditions in very strong gravitational and electromagnetic fields
near compact relativistic objects.

In particular, the
studies of phenomena in the vicinity of a neutron star make it
possible to obtain information on the properties of  matter
in the states, which are unattainable in the ground
laboratories. In this sense, a problem of a great
importance is to search and study the vacuum nonlinear
electrodynamics effects, which can occur in very strong
magnetic fields of such objects as pulsars ($B\sim 10^{12}$
G) and magnetars ($B\sim 10^{15}$ G).

As it is well known the nonlinear electrodynamics of
physical vacuum during a long time has no experimental
confirmation, and was usually regarded as an abstract
theoretical model. But now its status has changed
drastically. The last experiments on light-to-light
scattering made at Stanford (Burke et al., 1997) show that
electrodynamics in vacuum is really a nonlinear theory. Thus, the
different models of nonlinear electrodynamics of the vacuum
and their predictions (Ginzburg 1987; Alexandrov et al. 1989; Rosanov 1993;
Bakalov, D. et al. 1998, Denisov 2000a,b; Rikken $\&$ Rizzo 2000,
Denisov $\&$ Denisova 2001b,c) should be seriously tested in
experiments. However, the magnetic fields ($B\sim 10^{6}$ G)
available in ground laboratories give no such
opportunity because theory predicts that the typical value of
magnetic field induction necessary for essential
manifestation of electrodynamics nonlinearity in vacuum is
$B\sim 4.41\cdot 10^{13}$ G.

Since magnetic fields of some pulsars can be characterized
by such magnitudes, and for magnetars can reach
much greater values, this permits us to conclude that nonlinear
effects of electrodynamics in vacuum should be most
pronounced in the vicinity of such astrophysical objects.
Near a magnetic neutron star electromagnetic emission
undergoes the nonlinear electrodynamics effects of strong
magnetic fields. As a result, electromagnetic rays are
bended, the emission fluxes are dispersed and their spectra
and polarization states  change. The presence of super
strong magnetic field furthers the forming of a quite
extended magnetosphere with the radius of about several radii of a
neutron star. This magnetosphere is opaque, as a rule, for
the low-frequency part of electromagnetic spectrum. It will
be transparent only for X-rays and gamma rays.
Thus the spectrum, polarization and other parameters of
the detected hard emission from magnetic neutron stars are the only
data, which can give us the information on the main
regularities of the nonlinear electrodynamic interaction of
such emission with the strong magnetic field of a pulsar or
magnetar.

Thus, the most appropriate astrophysical objects where the
nonlinear electrodynamics effects can be manifested more
clearly are some types of rotation-powered pulsars,
accretion-powered pulsars and magnetars.

Different nonlinear electrodynamic effects in the vicinity of a
strongly magnetized neutron star were previously studied in the context of
quantum electrodynamics. In particular, vacuum birefringence and its effect
on the emitted spectra and on the propagation of photons in the neutron star
magnetosphere was discussed in the book of  Meszaros (1992). It was
found that vacuum effects dominate the polarization properties of the
normal modes of the near-neutron star medium. This gives rise to a
significant change in the medium opacity, thus the polarization
properties and transport of X-ray radiation from a neutron star's
magnetosphere can be altered by the magnetic vacuum effects (Meszaros
$\&$ Ventura 1978; Meszaros $\&$ Ventura 1979; B$\ddot o$rner $\&$
Meszaros 1979;
Meszaros et al. 1980; Meszaros $\&$ Bonazzola 1981, Denisov et al. 2002).
It was also shown that magnetic vacuum effects can change the spectra
of emitted radiation which leads to a signature in the spectra of
x-ray pulsars (Ventura et al. 1979). The nonlinear quantum electrodynamic
effects induced by an nonhomogeneous and non-stationary  magnetic field of
a neutron star, including the light bending in the plane of the magnetic
dipole equator, photon pair production and the frequency doubling and modulation
at the scattering of low frequency electromagnetic waves by the magnetic
field of an inclined rotator were discussed by Gal'tsov and Nikitina, 1983.
The quantum electrodynamic
effects in the accreting neutron stars, in particular, one and
two-photon Compton scattering in strong magnetic field and its effect
on the radiation processes (Bussard et al. 1986) as well as the vacuum
polarization effects in the field of a charged compact object
(De Lorenci et al., 2001) were also studied. However,
the analysis presented above concentrated on the validity of quantum
electrodynamics without comparison with possible alternative theories. Thus,
we try to obtain some specific predictions by using post-Maxwellian items of
different nonlinear generalizations of electrodynamics and then compare
the predictions of different theories with the goal to use the nonlinear
electrodynamics effects in neutron stars as a test for post-
Maxwellian effects.


\section{ Nonlinear models of vacuum electrodynamics}

As it is well known,  Maxwell electrodynamics is a
linear theory in the absence of matter. Its predictions
concerning a very wide field of problems (except the subatomic
level) are constantly confirming with better and better accuracy.
Quantum electrodynamics, which is based on Maxwell
electrodynamics complemented by the renormalization procedure,
also describes with good accuracy the various subatomic
processes, and, according to common opinion, is one of the best
physical theories.

However, some fundamental physical reasons indicate that
Maxwell electrodynamics is only the first approximation of more
general nonlinear vacuum electrodynamics, which can be
applied in the limit of weak electromagnetic fields.

The  electromagnetic field equations, which can be
obtained  using the Lagrange formalism, in any nonlinear
model of vacuum electrodynamics are equal to:
\begin{equation}
curl\ {\bf   H} ={1 \over c} {\partial {\bf  D} \over
\partial t},\ \ \ \
div \ {\bf D} =0,
\end{equation}
$$curl \ {\bf E} =-{1 \over c} {\partial {\bf  B} \over
\partial t},\ \ \ \ div \ {\bf B} =0.$$
However, in these equations the vectors $\bf  D$ and $\bf H$
depend on vectors $\bf  B$ and ${\bf   E},$ differently in
various models, since they are defined by the different
dependencies for the Lagrangian $L=L({\bf B},\ {\bf E})$:
\begin{equation}
{\bf  D}=4\pi {\partial L\over \partial {\bf E}},\ \
{\bf H}=-4\pi {\partial L\over \partial {\bf B}}.
\end{equation}

At the present time several nonlinear generalizations of the
Maxwell equations in vacuum are considered in the framework of
the field theory. The most well known among them are the Born-Infeld (BI)
electrodynamics (Born $\&$ Infeld 1934) and
the Heisenberg-Euler (HE) electrodynamics
 (Heisenberg $\&$ Euler 1936). These theories are based on absolutely
different principles, and, as a result they lead to
different electromagnetic field equations.

\subsection{ The Born-Infeld nonlinear electrodynamics}

Born and Infeld in their
research proceeded from the idea of a limited value of the
electromagnetic field energy of a point particle. This and
some other reasons led them to the following Lagrangian of
the nonlinear electrodynamics in vacuum:
\begin{equation}
L=-{1\over 4\pi a^2}\Big[\sqrt{1+a^2({\bf B}^2-{\bf E}^2)-
a^4({\bf B\cdot  E})^2}-1\Big],
\end{equation}
where $a$ - is the constant with units reciprocal to
the units of magnetic field induction.

Born and Infeld defined this constant from the assumption
that the origin of all rest energy of an electron ${\cal
E}_0$ is electromagnetic. As a result, the estimate
 $1/a= 9.18\cdot 10^{15} $ G was obtained (Born $\&$ Infeld 1934)
from the atomic physics constraints.

Thus, although this theory has a definite Lagrangian, it
is to a large extent phenomenological, and to verify it is
necessary, first of all, to measure experimentally the value of $a^2$
parameter or  at least to estimate its upper limit.

In view of relations (2) and (3), in the BI nonlinear
electrodynamics vectors $\bf D$ and ${\bf H},$ are the
following functions of vectors
$\bf  B$ and ${\bf   E}:$
$${\bf  D}=
{{\bf  E}+a^2({\bf B\cdot  E}){\bf B}\over \sqrt{1+a^2({\bf B}^2-
{\bf E}^2)-
a^4({\bf B\cdot   E})^2}},$$
$${\bf   H}=
{{\bf B}-a^2({\bf B\cdot  E}){\bf E}\over\sqrt{1+a^2({\bf  B}^2-
{\bf  E}^2)-a^4({\bf B\cdot E})^2}}.$$
For the field magnitudes, which can be achieved in the ground
laboratories, the values $a^2{\bf E}^2$ and $a^2{\bf B}^2$
are much less than one. In this case Lagrangian of the BI
nonlinear electrodynamics can be expanded into the
small parameters
$a^2{\bf E}^2<<1$ and  $a^2{\bf B}^2<<1:$
\begin{equation}
L=-{1\over 8\pi }({\bf B}^2-{\bf E}^2)+{a^2\over 32\pi
}\Big[({\bf B}^2-{\bf E}^2)^2+4({\bf B\cdot E})^2\Big].
\end{equation}
The first term of this expansion is the Lagrangian of the
Maxwell electrodynamics, and the other term is the non
linear correction to it, which is proportional to the above mentioned
 small parameters.

It was shown by Cecotti $\&$ Ferrara (1987),
Denisov (2000a), Denisov et al. (2000), that the BI
electrodynamics has a number of very interesting properties
and in many ways it is remarkable theory.

First, as it was mentioned above, the energy of the electromagnetic field of
a point charge is a finite quantity in the framework of this theory.

Second, the ideology of this theory is very close to Einstein's
idea of introducing a non-symmetric metric tensor $G_{ik}\neq
G_{ki}$ with the symmetric part corresponding to the usual metric
tensor $g_{ik}$ and the antisymmetric part, corresponding to the
electromagnetic field tensor $F_{ik}:$
$G_{ik}=g_{ik}+aF_{ik}.$
Using the relations of tensor algebra (Denisova and Mehta 1997)
 it is not difficult to show
that  the Lagrangian (3) can be written as
$$L=-{1\over 4\pi a^2}[\sqrt{-G}-\sqrt{-g}].$$

Besides,
though the velocity of an electromagnetic
wave depends on the values of the fields $B^2$ and $E^2$ in this
theory, it does not exceed the speed of light  $c$ in Maxwell's
electrodynamics.
It should be noted also that BI electrodynamics can be
obtained from more general sypersymmetric theories.

Thus, the BI electrodynamics in many respects constitutes a
distinguished theory. However, to the present time this theory is not
developed enough because of the lack of quantitive estimations of
different effects. In particular, none of the scientific
publications contain any calculations in the BI theory, which could
give estimates of the
probability of $e^-e^+$ pair production in the SLAC experiments
 (Burke et al., 1997).
On the other hand, it is necessary to note, that to the present
time there are no experiments, which would rejected this
theory.

\subsection{ The Heisenberg-Euler nonlinear electrodynamics}

As it is well known the HE nonlinear
electrodynamics is based on the quantum electrodynamic (QED)
effect of electron-positron vacuum polarization by
electromagnetic fields.

Thus, the linear Maxwell electrodynamics is only the first
approximation of a more general nonlinear electrodynamics (in
 vacuum), which can be used in the case of weak
electromagnetic fields, when its magnitudes $\bf B$ and $\bf E$
are much smaller than the characteristic quantum
electrodynamic value $B_q=m^2c^3/e\hbar =4.41\cdot
10^{13}$ G,
where $m$ is the mass of an electron, $e$ is the module of
its charge, $\hbar $ - is the Plank constant.

The accurate form of the Lagrangian in this theory has not been defined yet.
However, for  "weak" electromagnetic fields
corrections to the Maxwell Lagrangian in the first
non-vanishing order of the quantum electrodynamics perturbation
theory have a strictly  defined form. As it can be seen
from calculations (Heisenberg $\&$ Euler 1936), if electromagnetic fields
are not
strong $(B<<B_q,\ E<<B_q)$ the first two terms in the vacuum
electromagnetic field nonlinear Lagrangian expansion in
the small parameters $({\bf B}^2- {\bf E}^2)/
B_q^2$ and $({\bf B\cdot E})/B_q^2,$ should have the form:
\begin{equation}
L=-{1\over 8 \pi }[{\bf B}^2-{\bf E}^2]+{\alpha \over
360\pi^2B^2_q}
\Big\{({\bf B}^2-{\bf E}^2)^2+7({\bf B\cdot E})^2\Big\},
\end{equation}
where $\alpha =e^2/\hbar c\approx 1/137$ is the fine
structure constant.

The vectors $\bf D$ and $\bf H$, which are contained in
equations (1), are also nonlinear functions of the vectors $\bf B$
and $\bf E$ in this theory. In the first non-vanishing
approximation of quantum electrodynamics their form is:
$${\bf D}=
{\bf E}+{\alpha \over 45 \pi B^2_q} \Big\{2({\bf E}^2-{\bf B}^2)
{\bf E}+7({\bf B\cdot  E}){\bf B}\Big\},$$
$${\bf H}={\bf B}+{\alpha \over 45 \pi B^2_q}
\Big\{2({\bf E}^2-{\bf B}^2){\bf B}-7({\bf B\cdot E}){\bf E}\Big\}.$$
Comparing expressions (4) and (5), it is easy to see
that they can not be reduced to each other by  any choice
of  the $a^2$ constant. This means, that nonlinear
electrodynamics with Lagrangians  (3) and (5) are essentially
different theories. Thus,  experimental verification of predictions of these
theories  as well as the solving of problem of
their adequacy to reality are of great interest.

\subsection{{\bf Other} models of nonlinear electrodynamics}

Other models of nonlinear vacuum electrodynamics
are also discussed in the field theory. It is quite
natural, that in other theoretical models of nonlinear
electrodynamics the coefficients at the terms
$({\bf B}^2-{\bf E}^2)^2$ and $({\bf B\cdot E})^2$ in the
Lagrangian expansion   can be absolutely arbitrary.

Thus, to choose that nonlinear electrodynamics, which is most
adequate to  nature, it is necessary to calculate the nonlinear
effects in different theories and to compare their
predictions with the results of the corresponding experiments.

To make such calculations easier in the approximation of a
weak electromagnetic field, we will use a parameterized post-Maxwell
formalism, which was elaborated by  Denisov $\&$ Denisova (2001a,c).
This
formalism is similar, in some sense, to the parameterized
post-Newton formalism in the theory of gravitation (Will 1981),
which is commonly used  for calculating different
gravitational effects in the weak field of the Solar system.

We will assume, that the main prerequisite for this formalism is
that the Lagrangian of nonlinear electrodynamics in
vacuum is an analytical function of invariants $J_1 = ({\bf E}^2
- {\bf B}^2) / B^2_q$ and $J_2 = ({\bf E\cdot B})^2 / B^4_q,$ at
least, near their zero values. Thus, in the case of a weak
electromagnetic field $(J_1<<1,\ J_2<<1)$ this Lagrangian can
be expanded into a converging set in integer powers of
these invariants:
\begin{equation}
L={B_q^2\over 8\pi}\sum\limits_{n=0}^\infty
\sum\limits_{m=0}^\infty L_{nm}J_1^nJ_2^m.
\end{equation}
Since, at $J_1\to  0, J_2\to  0$ the theory with
Lagrangian (6) should be reduced to  Maxwell
electrodynamics, then $L_{00} = 0, L_{10} = 1$.

For such an approach, a quite definite number of post-Maxwell
parameters $L_{nm}$ will correspond to each nonlinear
electrodynamics. From the point of view of the experiments
in a weak electromagnetic field, we can conclude, that one nonlinear
electrodynamics will differ from the other only
by the values of these parameters.

If we limit oneself only to a few first terms in  expansion
(6), then according to the parameterized post-Maxwell
formalism the generalized Lagrangian of the nonlinear
vacuum electrodynamics in the case of weak fields can be
represented as (Denisov $\&$ Denisova 2001a,c):
\begin{equation}
L={1\over 8\pi }\Big\{[{\bf  E}^2-{\bf  B}^2]+
\xi [\eta_1({\bf  E}^2-{\bf  B}^2)^2+
4\eta_2({\bf  B\cdot   E})^2]\Big\},
\end{equation}
where $\xi =1/B^2_q,$ and the value of the dimension-less post-Maxwell
parameters $\eta_1$ and $\eta_2$ depend on the
choice of the model of nonlinear vacuum electrodynamics.

In particular, in the nonlinear HE
electrodynamics parameters $\eta_1$ and $\eta_2$
have  quite definite values
$\eta_1=\alpha /(45 \pi )=5.1\cdot 10^{-5}, \ \eta_2=7\alpha
/(180 \pi )=9.0\cdot 10^{-5},$ while in the BI theory they
can be expressed through the same unknown
constant $a^2:$ $\eta_1=\eta_2=a^2B_q^2/4.$

Substituting Lagrangian (7) into expressions (2), we obtain
vectors $\bf D$ and $\bf H$ of the parameterized nonlinear vacuum
electrodynamics:
$${\bf D}=
{\bf E}+2\xi \Big\{\eta_1({\bf E}^2-{\bf B}^2){\bf E}+2\eta_2
({\bf B\cdot  E}){\bf B}\Big\},$$
$${\bf H}={\bf B}+2\xi \Big\{\eta_1({\bf E}^2-{\bf B}^2){\bf B}-
2\eta_2({\bf B\cdot  E}){\bf E}\Big\}.$$
Thus, the post-Maxwell formalism without focusing on the details
of one or another nonlinear electrodynamics, its equations,
hypotheses and postulates, on all its theoretical
composition takes into account only the final result:
expansion of Lagrangian, which according to the given theory is
valid in the weak electromagnetic field approach. Further
analysis of the theories and revealing of the concordance of
their predictions with experimental results is quite general
and can be reduced to obtaining the answers to two questions: what are
the values of post-Maxwell parameters in the studied theory and
what are the parameter values according to the
results of corresponding experiments.

Thus, one of the goals of this formalism is the calculation
of "weak" nonlinear electrodynamic effects disrespectively
to any nonlinear theory. The goals of the theory and
the experiment in this case should be not only the search for
such an effect (which can refute one or another nonlinear
electrodynamics), but also experiments with
the purposes of determination (with necessary accuracy) of the
all post-Maxwell parameter values.

\section{ The effect of nonlinear-electrodynamic bending of a ray}

The electromagnetic emission is the main channel carrying
information on nonlinear electrodynamic effects,
which can occur in the magnetic dipole field of
astrophysical objects. The electromagnetic ray is exactly the
agent, which passing through the neutron star magnetic field
undergoes nonlinear electromagnetic influence from this
field independently of the spectral range. Studying the main
parameters of incoming electromagnetic emission, such as
dependence of a ray bending angle on impact distance, the
law of emission intensity decreasing in the course of time,
etc., it is possible  (Denisov, V.I. et al. 2001)
to reveal the main dependencies of nonlinear electrodynamic
interactions of electromagnetic fields.

\subsection{ Dispersion Equation}

In order to study the laws of weak electromagnetic waves propagation
in the dipole magnetic field of a neutron star, we will
obtain the dispersion equation. We will assume that a "weak"
plane electromagnetic wave propagates through the permanent
magnetic field $\bf{B}_0$  of a neutron sta. Then in the
geometric optics approach we can write the following
relations:
\begin{equation}
{\bf E}={\bf e}\ \exp[-i(\omega t-{\bf k\cdot r})],
\end{equation}
$${\bf B}={\bf B}_0+{\bf b}\ \exp[-i(\omega t-{\bf k\cdot
r})],$$
where $\omega $ is the frequency, $\bf k$ is the wave
vector and vectors
$\bf b$ and $\bf e$ are slowly changing functions of $t$ and
${\bf r},$ in comparison with the
$\exp[i(\omega t-{\bf k\cdot r})]$ function.

Under this approach the dispersion equation can be obtained
from Lagrangian (7) directly. However, though the obtained result will
be true, quite legitimate questions about its correctness
will arise during calculations. Thus, to ensure the
necessary accuracy of calculations, we will add to Lagrangian
(7) the terms of higher approximations and, hence,  write
it with surplus accuracy:
\begin{equation}
L={1\over 8\pi }\Big\{[{\bf E}^2-{\bf B}^2]+
\xi [\eta_1({\bf E}^2-{\bf B}^2)^2+
4\eta_2({\bf B\cdot E})^2]+
\end{equation}
$$+\xi^2[\eta_3({\bf E}^2-{\bf B}^2)^3+
\eta_4({\bf E}^2-{\bf B}^2)({\bf B\cdot E})^2]\Big\}.$$
This relation is used to obtain the dispersion
equation. The details of the transitions are presented in
Appendix A. The result is that according to equations of the
post-Maxwell nonlinear vacuum electrodynamics in the
presence of a permanent and regular magnetic field "weak", generally,
plane electromagnetic waves of two types can
propagate in any direction. The corresponding dispersion
equations are:
\begin{equation}
\omega_1({\bf k})=ck\Big\{1-{2\eta_1 \xi \over k^2}[{\bf k\times
 B}_0]^2+O(\xi^2{\bf B}_0^4)\Big\},
\end{equation}
$$\omega_2({\bf k})=ck\Big\{1-{2\eta_2 \xi \over k^2}
[{\bf k\times B}_0]^2+O(\xi^2{\bf B}_0^4)\Big\}.$$
It is necessary to note, that the same dispersion equations
can also be obtained from a simpler Lagrangian (7).

As it was shown previously (Denisov 2000a), the exact
dispersion equation for electromagnetic wave propagating in the
magnetic field ${\bf B}_0$ in the BI theory has the form
\begin{equation}
(1+a^2{\bf B}_0^2){\omega^2\over c^2}-{\bf k}^2-a^2({\bf k}\ {\bf B}_0)^2=0,
\end{equation}
independently of its polarization at the any $a^2{\bf B}_0^2$ value.

The solution of Maxwell equations (1) for electromagnetic
waves propagating in the magnetic field shows that at
$\eta_1\neq \eta_2$ the waves of both types with the
dispersion equations (10) are polarized linearly in  mutually
normal planes and propagate with different group velocities. This
property of electromagnetic waves is well known as birefringence.

At $\eta_1=\eta_2$ the both types of electromagnetic waves
will coincide to the accuracy of terms proportional to
$\xi^2$. As a result, electromagnetic waves of the same type
with arbitrary polarization will propagate in each direction.

Let us now find now the eikonal equation for an electromagnetic
wave propagating in the dipole magnetic field of a neutron
star under the laws of nonlinear vacuum electrodynamics.
For this purpose we will raise relations (10) to the second power.
Retaining terms linear in $\xi $ and taking into account that
$\omega =\partial S/\partial t, \ {\bf k}={\bf \nabla} S,$
we obtain:
\begin{equation}
{1\over c^2}\Big({\partial S\over \partial t}\Big)^2-
\Big[1-4\eta_1\xi {\bf B}_0^2\Big]\Big({\bf \nabla} S\Big)^2-
4\eta_1\xi \Big({\bf B}_0{\bf\cdot} {\bf \nabla} S\Big)^2=0,
\end{equation}
$${1\over c^2}\Big({\partial S\over \partial t}\Big)^2-
\Big[1-4\eta_2\xi {\bf B}_0^2\Big]\Big({\bf \nabla} S\Big)^2-
4\eta_2\xi \Big({\bf B}_0{\bf\cdot }{\bf \nabla }S\Big)^2=0.$$
In the BI theory, as it follows from the relation (11),
the eikonal equation valid for any $a^2{\bf B}_0^2,$ values has the form
\begin{equation}
{1\over c^2}\Big({\partial S\over \partial t}\Big)^2[1+a^2{\bf B}_0^2]-
({\bf \nabla }S)^2-a^2({\bf \nabla }S{\bf B}_0)^2=0.
\end{equation}
The solution of these equations in common case is not known.
Thus, further on we will consider solution of the equations (12)
only for the rays laying in the dipole magnetic field
equator plane.
Determination of the evident form of eikonal equations
(12) for electromagnetic wave propagating in the field of
a magnetic dipole permits us to study the main nonlinear
electrodynamic effects, which should occur in the magnetic
fields of pulsars and magnetars.

It is necessary to note, that the effects discussed below can partially be
caused by the gravitational field of  neutron
stars. However, in the first order on perturbation the
gravitational and nonlinear electrodynamic parts are
additive. Since the gravitational effects were studied
repeatedly (Epstein $\&$ Shapiro 1980; Will 1981; Meszaros $\&$ Riffert
1988; Riffert $\&$ Meszaros 1988)  and to the present time studies of
gravitaional lensing effects in astrophysics remains the main goal of
some observational programs (Sutherland et al. 1996),  we will not
consider them here and will pay attention mainly to the effects of
nonlinear vacuum electrodynamics.

\subsection{The Bending of a ray from a source located at a limited
distance from a neutron star in its magnetic field}

Let us  denote the plane normal to the magnetic dipole
momentum vector $\bf m$, as plane $XOY.$ In this case only
one component of the vector ${\bf m}=|{\bf m}|{\bf e}_z$  will be
nonzero and vector ${\bf B}_0$ in this plane can be represented
as:
${\bf B}_0=-|{\bf m}|{\bf e}_z/r^3.$

Hence the first of the eikonal equation (12) for electromagnetic wave
polarized in the $XOY$ plane, which ray lay in the same plane, will be:
\begin{equation}
{1\over c^2}\Big({\partial S_1\over \partial t}\Big)^2-
\Big[1-{4\eta_1\xi {\bf m}^2\over r^6}\Big]
\Big[\Big({\partial S_1\over \partial r}\Big)^2+{1\over r^2}
\Big({\partial S_1\over \partial \varphi
}\Big)^2\Big]=0.
\end{equation}
A similar equation with the replacement of the $\eta_1$ parameter by the
$\eta_2$ parameter and  $S_1$ by $S_2,$ can be written
for a ray of electromagnetic wave polarized along the  $z$ axis.

As it is accepted in theoretical mechanics (Landau $\&$ Lifshitz 1984),
we will  find the partial solution of equation (14) using  the
variables separation method. As a result, we obtain:
\begin{equation}
S_1=-{\cal E}_0t+\alpha \varphi +\int\limits^rdr
\sqrt{{{\cal E}_0^2\over c^2}\Big[1+{4\eta_1\xi {\bf m}^2
\over r^6}\Big]-{\alpha^2\over r^2}},
\end{equation}
where ${\cal E}_0,\ \alpha $ are the constants of
integration and all calculations were made with accuracy,
linear in the small value $\eta_1\xi {\bf m}^2/ r^6.$

It should be noted, that in the magnetic equator plane the
expression (15) is also the solution of BI exact eikonal
equation (13), if we take into account, that in this theory
$\eta_1=\eta_2=a^2B_q^2/4.$

Using  relation (15) we can determine the kinematic and dynamic parameters
describing   photon propagation in the dipole magnetic field.

Let us consider the case, when the
gamma ray source is located at a limited distance $l_1$
from a neutron star or even in its nearest vicinity. This
can take place in the case of an accretion-powered
pulsar. The latter conditions can be realized in the case of a
rotation-powered gamma-
pulsar (if the polar cap models are valid).

Let us denote the distance from neutron star to the detector as $l_2.$
Then the
distance $l_1$ is much smaller than $l_2$ and comparable with
the neutron star radius $R.$
Hence, the dependence of
impact distance on time for a circular orbit in the first
approximation can be represented as: $b(t)=b_0+R_1\cos\Omega t,$
where $R_1$ is the orbit radius, $\Omega $ is the orbital
frequency.

If we are  considering the propagation of a X-ray or
gamma ray photon from a source located near a
galactic neutron star, it is convenient to direct the $X$
and $Y$ axes in such a way, that a ray from the
source travels along the $X$ axis with the impact distance
$b$, the center of the dipole magnetic field is placed in the
center of the coordinate system (see Fig. 1) and the spacecraft
with the detectors is located at the distance $l_2$ from the
center of the coordinate system near the point $x\approx  l_2$.

   \begin{figure*}
   \centering
   \includegraphics{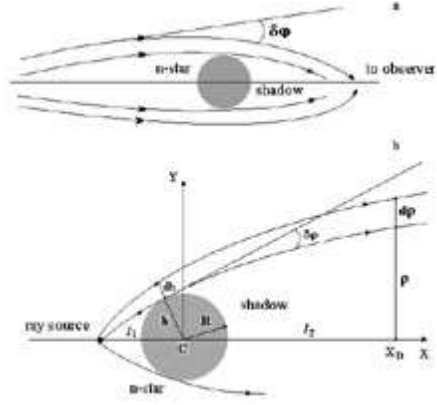}
   \caption{The ray bending and shadow region in the case
$l_1>>R$ (a); the ray bending and shadow region in the case
$l_1\geq R$ (b). } 
    \end{figure*}
Since the value $\xi {\bf m}^2/r^6$ is small, in order to find the
bending angles $\delta\varphi^*_{1,2}$
we can use the algorithm (Darwin 1961), well established for calculations of
angles of light gravitational bending.
The details of calculations of the bending angle $\delta\varphi^*_1$
in the case of an electromagnetic wave polarized in the $XOY$ plane,
and $\delta\varphi^*_2$ for an electromagnetic wave polarized
along the $z$ axis, are presented in Appendix B. For the case, when
$b<l_1<<l_2,$ and $l_2>>b$ we obtained following
relations for the bending angles:
\begin{equation}
\delta\varphi_1^*=\arctan\left[{bQ_1\over \sqrt{Q_1^2l_1^2-1}}\right]
+{15\pi \eta_1\xi {\bf  m}^2\over 4 b^6}-
\end{equation}
$$-\big[1+{15 \eta_1\xi {\bf  m}^2\over 4 b^6}\big]\arcsin({1\over Q_1l_1})-
$$
$$-{\eta_1\xi {\bf  m}^2\over 16 b^6}\big\{\sin[4\arcsin({1\over Q_1l_1})]
-16\sin[2\arcsin({1\over Q_1l_1})]\big\},$$
$$\delta\varphi_2^*=\arctan\left[{bQ_2\over \sqrt{Q_2^2l_1^2-1}}\right]
+{15\pi \eta_2\xi {\bf  m}^2\over 4 b^6}-$$
$$-\big[1+{15 \eta_2\xi {\bf  m}^2\over 4 b^6}\big]\arcsin({1\over
Q_2l_1})-$$
$$-{\eta_2\xi {\bf  m}^2\over 16 b^6}\big\{\sin[4\arcsin({1\over
Q_2l_1})] -16\sin[2\arcsin({1\over Q_2l_1})]\big\},$$
where
$$Q_1={1\over b}\big[1+{2\eta_1\xi {\bf m}^2\over b^6}\big],\ \ \ \
Q_2={1\over b}\big[1+{2\eta_2\xi {\bf m}^2\over b^6}\big].$$
The plus sign  in this relation shows, that the magnetic dipole
field in the magnetic equator plane effects
the electromagnetic waves as a convex lens. To illustrate
dependeces (16) in BI and HE theories we plot $\delta\varphi_1^*$ and
$\delta\varphi_2^*$ versus $b/l_1$ for different values $B_0/B_q$ (Fig. 2
-- Fig. 4).

   \begin{figure*}
   \centering
   \includegraphics{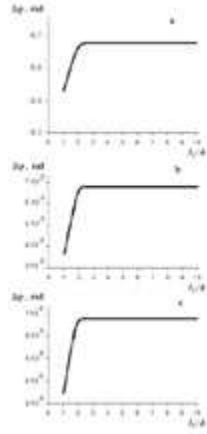}
   \caption{The BI dependence
              $\delta\varphi $ versus $l_1/b$ for different values 
${\bf B}^2(b)/B_q^2:$  $(a)\ {\bf B}^2(b)/B_q^2 =10^{-2},\ (b)\ 
{\bf B}^2(b)/B_q^2=10,
(c)\ {\bf B}^2(b)/B_q^2=1$ ($ {\bf B}^2(b)$ is the square of magnetic field at 
the distance $b$
from the neutron star, $\delta\varphi_1^*=\delta\varphi_2^*).$}
     \end{figure*}

   \begin{figure*}
   \centering
   \includegraphics{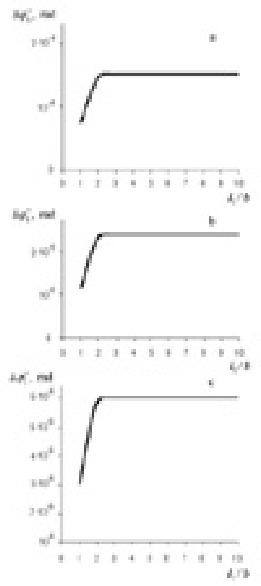}
   \caption{The HE dependence
              $\delta\varphi_1^* $ versus $l_1/b$ for different values 
${\bf B}^2(b)/B_q^2:$  $(a)\ {\bf B}^2(b)/B_q^2 =0.25,\ (b)\ 
{\bf B}^2(b)/B_q^2=0.04,
(c)\ {\bf B}^2(b)/B_q^2=0.01$ ($ {\bf B}^2(b)$ is the square of 
magnetic field at the distance $b$ from the neutron star).}

    \end{figure*}

%
   \begin{figure*}
   \centering
   \includegraphics{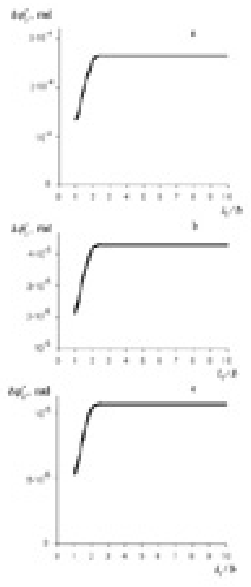}
   \caption{The HE dependence
              $\delta\varphi_2^* $ versus $l_1/b$ for different values 
${\bf B}^2(b)/B_q^2:$  $(a)\ {\bf B}^2(b)/B_q^2 =0.25,\ (b)\ 
{\bf B}^2(b)/B_q^2=0.04,
(c)\ {\bf B}^2(b)/B_q^2=0.01$ ($ {\bf B}^2(b)$ is the square of 
magnetic field at the distance $b$ from the neutron star).}

    \end{figure*}

Thus, nonlinear models of vacuum  electrodynamics
with $\eta_1\neq \eta_2,$ predict different angles of  ray
bending for electromagnetic waves with different polarization.

It should be noted, that besides the nonlinear electrodynamic bending
(16) of electromagnetic rays will also undergo the well known
gravitational bending. However, because of the different bending
angle dependence on the impact distance $b$ ($1/b$ and $1/b^2$ in the
case of gravitational bending (Epstein $\&$ Shapiro 1980; Will 1981;
Meszaros $\&$ Riffert 1988; Riffert $\&$ Meszaros 1988) and $1/b^6$ in
the case of nonlinear electrodynamic bending) mathematical processing
allows to resolve each of these parts from the observational data if
the time  dependence of impact distance $b$ is harmonical
$b(t)=b_0+R_1\cos\Omega t.$

\section{ The effect of gamma-ray flux scattering in the neutron
star magnetic field}

The nonlinear electrodynamic bending of rays in the magnetic dipole field
differs significantly from gravitation bending of these rays. As it is
well-known, the ray bending in the Schwarzschild gravitation field occurs
in the plane containing this ray and the star center. Thus, gravitation ray
bending leads to the increase of electromagnetic flux detected by the
instrument located on the other side of gravitation center.

In the general case the nonlinear electrodynamic bending of electromagnetic
ray is not planar. It can be discribed as the bending on the series of
magnetic force lines, when a ray undergoes bending on every force line in
the plane containing the normal to the surface $[{\bf B}_0\ {\bf k}]^2/k^2=
const.$ As a
result the nonlinear interaction of electromagnetic wave with the dipole
magnetic field leads in the general case to the appearing of the ray
curvature and twisting.

For a complete and detailed study of the principles of nonlinear
electrodynamic lensing, it is necessary to solve Eqs. (12) in the very
general case, but not only for rays laying in the neutron star magnetic
equator plane. However, to the present time the solution of these equations
is unknown.  Formulas (16) allow us to study only one particular case
of nonlinear electrodynamic lensing, when the ray source is near the
neutron star, at the distance $l_1\sim R$, and electromagnetic
emission is a bunch of rays laying in the small vicinity of magnetic
equator plane.

When studying gravitation lensing of light emission, it can usually be
assumed that the distance between light source and the gravitating star
$l_1$ is larger than the star radius $R$. In this case ray bending leads to
the abrupt cutting of the field of shadow behind the opaque star (see Fig.
1a).  As a result, the light flux detected at the large distance will
be much larger than in the case of gravitating center absence (the
effect of gravitational lensing).

The sources of gamma-rays are located mainly not far, or according to
some models (Harding, 2000), 
even near the pulsar's or magnetar's surface: $l_1\sim R$.
Thus, even with account for the gravitational and nonlinear electromagnetic
bending of gamma-rays, we should assume the existence of some shadow cone
behind the neutron star, where gamma-rays never penetrate (see Fig.
1b).

The main condition of the existence of such cone is the non-equality:
$\delta\varphi_{max} < \arcsin(R/l_1).$ Let us consider this case in
details.

We will assume, that a magnetized neutron star is located in the coordinate
center, the gamma ray source is placed in the $XOY$ plane at the distance
of $l_1$ from the neutron star (left panel of Fig. 1b) and the
spacecraft carrying the gamma ray detector is located at a large
distance from the star, at the point $x_D=l_2>>l_1,\ y_D=\rho .$ Let
us calculate the value of gamma-ray flux $I_{out}$ detected by the
presence of gravitation and nonlinear electrodynamic bending of rays.

Let us assume, that in the case of the absence of such bending, the
detector will measure the energy flux $I_0=A/\sigma [\rho^2+(l_1+l_2)^2],$
where $\sigma $ is the value of gamma-ray beam solid angle, $A$ is some
constant characterizing the gamma-ray source intensity.

It is quite appropriate to chose the sector of a ring of radius, $b,$
width, $db$ and angle opening $\delta\theta <<1$ as the input aperture.
This aperture area will be equal to:  $S_{in}=bdb\delta\theta .$

In the input aperture plane the gamma-ray flux $I_{in}$ will be:
$I_{in}=A/\sigma [l_1^2-R^2]=I_0[\rho^2+(l_1+l_2)^2]/[l_1^2-R^2].$ Since in
the field of the magnetic equator it is possible to consider the
gravitational and nonlinear electrodynamic bending of rays as occurring
approximately in the planes containing the neutron star center, the output
aperture can also be presented as the sector of a ring of radius, $\rho $,
width, $d\rho .$, and angle opening $\delta\theta .$ Thus the output
aperture area will be equal to:  $S_{out}=\rho d\rho \delta\theta .$

Since the energy value transferred per unit time along the bunch of rays
does not depend on the distance, then from the equality
$I_{in}S_{in}=I_{out}S_{out}$ we obtain:
\begin{equation}
I_{out}={I_0[\rho^2+(l_1+l_2)^2]\over (l_1^2-R^2)}{b\over \rho} {db\over
d\rho}. 
\end{equation}
In this relation values $\rho$ and  $b$ are not independent, but
are connected by the equation, which can easily be obtained from
geometrical consideration (see Fig. 1b):
\begin{equation}
\rho=b\cos\alpha_0+\big[b\sin \alpha_0+l_2\big]
tg\Big(\alpha_0-{2r_g\over b}-\delta\varphi_{1,2}^*\Big) ,
\end{equation}
 where
$\alpha_0=\arcsin(b/l_1),$ $r_g$ is the Schwarzschild radius of
neutron star, and $\delta\varphi^*_{1,2}$ is defined by espression
(16).

This equation is transcendental relatively to the impact parameter $b.$ Its
analytical approximate solution can be obtained only under a large number
of restrictions on the parameters contained in it.  In particular, if
$R\sim 10^2$ km, $l_1\sim 10^4$ km, $r_g\sim 0.2$ km, $l_2\sim 10^{17}$km,
 $\delta\varphi^*_{1,2}<10^{-2},$ then is transcendental equation (18) by
$2R\geq b\geq R$ can be approximate by algebraic equation of the 7th order
relatively to the impact parameter $b:$ 
$$\rho=l_2\Big[{b\over l_1}-
{2r_g\over b}-{15\pi\xi\eta_{1,2}|{\bf m}|^2 \over 4b^6}\Big].$$ 
Since $b\geq R,$
and $R/l_1\sim\ 10^{-2},$ the last two terms in the squared brackets give
the small correction to $b/l_1,$ which decreases with the increasing $b.$
As a result, the approximate solution of this equation can be written as:
\begin{equation}
b={l_1\rho\over l_2}\Big\{1+{l_2\over \rho}\Big[{2r_gl_2\over l_1\rho}
+{15\pi\xi\eta_{1,2}|{\bf m}|^2l_2^6\over 4l_1^6\rho^6}+ O\Big({r_g^2l_2^2\over
l_1^2\rho^2}\Big)\Big]\Big\},
\end{equation}
 and this solution is valid only by
variation of $\rho$ within the limits:
$2l_2R/l_1>\rho>l_2[R/l_1-2r_g/R-15\pi\xi\eta_{1,2}|{\bf m}|^2/(4R^6)].$

Substituting expression (19) in the equality (17), we then obtain, that in this
field of variation of $\rho$ the detected gamma-ray flux will be equal:
\begin{equation}
I_{out}=I_0\Big[1-{75\pi\xi\eta_{1,2}|{\bf m}|^2l_2^7\over 4l_1^6\rho^7}+
O\Big({r_g^2l_2^3\over l_1^2\rho^3}\Big)\Big]<I_0.
\end{equation}
Thus, gravitation and
nonlinear electrodynamic bending of gamma-rays leads to a decrease of the
detected flux in the considered part of space. It is caused by
consideration, that in the absence of the ray bending the illuminated area
$\rho\geq R l_2/l_1$ is less than illuminated area $\rho\geq
l_2[R/l_1-2r_g/R- 15\pi\xi\eta_{1,2}|{\bf m}|^2/(4R^6)]$ in the presence of
bending.  As a result of the ray bending, part of the gamma-ray energy flux
is transferred from the field of space $\rho\geq R l_2/l_1$ to the field of
space bounded by the conic surfaces  $\rho = R l_2/l_1$  and  $\rho=
l_2[R/l_1-2r_g/R- 15\pi\xi\eta_{1,2}|{\bf m}|^2/(4R^6)],$ thus decreasing the
flux of energy outside the cone $\rho\geq R l_2/l_1.$

We will make some estimates. If we use the above presented numerical values
of $R, l_1, r_g, l_2$ and if to assume that in the case of some gamma-ray
pulsars near the neutron star surface $B_0=2\cdot 10^{13}\ G,$ (Thompson
2000) we obtain, that relative value $(I_0-I_{out})/I_0$ changes from
$10^{-2}$ at $\rho= R l_2/l_1\sim 10^{15}$ km to $10^{-4}$ at $\rho= 2R
l_2/l_1\sim 2\cdot 10^{15}$ km.

Due to the larger magnetic field $B_0\sim\cdot 10^{15}-10^{16}\ G,$ (Zhang
$\&$ Harding 2000; Duncan $\&$ Thompson 1992; Thompson $\&$ Duncan
1995, 1996) in the case of a magnetar this value will change in wider
boundaries. However, the correct estimates in post-Maxwellian
approximation can be made only in the frame of the BI
electrodynamics, since in the HE theory, if the magnetic field is
greater than the $B_q$ value, the Lagrangian should be changed
drastically, because it sh ould contain logarithmic terms.

\section{ Analysis of x-ray and gamma-ray astronomy technique applications
for observing nonlinear electrodynamic effects }

The modern accuracy of gamma ray flux parameter measurements provided by
the use of extra-Terrestrial techniques is much worse than the accuracy of
measurements of similar parameters in the optical range (Boyarchuk et al.
1999).  However, due to the opacity of the strongly magnetized neutron star
magnetosphere to the optical emission, X-rays and gamma rays give the
unique possibility to research for nonlinear electrodynamic effects in the
strong magnetic field of a neutron star.
{\bf 
\subsection{The effect of gamma-ray flux dispersion}}

Taking into account the capabilities of modern X-ray and gamma ray space
observatories discussed in the Introduction, we will analyze, which of the
above mentioned effects can be observed using space astronomy techniques.

In the case, when the distance $l_1$ between the gamma ray source and the
neutron star is comparable with the neutron star radius, the energy flux
 (20) is close to one even for the impact
distances comparable to the star radius. Thus, the rays which underwent
significant nonlinear electromagnetic influence from the neutron star
magnetic field can be weakly dispersed, and as a result their intensity at
the point of observation will be sufficient for detection.

Let us estimate the magnitudes of these effects. However, first of all it
is necessary to make the following clarification. All the above discussed
effects depend not only on the magnetic field value, but also on the choice
of the model nonlinear vacuum electrodynamics (since transition from one
model to another changes the value of post-Maxwell parameters $\eta_1$ and
$\eta_2$). At present it is impossible to say, which value of these
parameters is agrees with observational data.

It is necessary to note, that in the HE nonlinear electrodynamics (Ritus
1986) the expansion parameter is $p= ({\bf B}/B_q)^2.$ By  $B=2\cdot 10^{13}$ G
this parameter is equal to  $p=0.2$ and post-Maxwellian expansion (5) of
QED still can be used for estimation of nonlinear effects values.

Thus, when making the estimates we will take into account the most well
known models of nonlinear vacuum electrodynamics, such as the BI
electrodynamics and the nonlinear electrodynamics, which is the direct
consequence of quantum electrodynamics. According to these theories the
post-Maxwell corrections to the Lagrangian of nonlinear electrodynamics are
about $\sim 10^{-4}{\bf B}^2/B_q^2.$

 However, it is necessary to note, that these estimates are quite
conditional, and should be considered as estimates of concrete theories. It
may be, that some other theory is more adequate to nature, and will give
other estimates of the studied nonlinear electrodynamic effect values in
the discussed experiments. During the space experiments it is necessary to
search for and measure record all the above mentioned effects in order to
obtain the values of post-Maxwell parameters $\eta_1$ and $\eta_2$ from the
results of observational data processing. Thus, the purpose of
astrophysical observations is not only the testing the predictions of one
or another nonlinear vacuum  electrodynamics model, but also what is more
important the measuring of parameters $\eta_1$ and $\eta_2$.

Using equalities (16), it is not difficult to obtain, that the maximum
value of the nonlinear electrodynamic bending angle of gamma-rays in the
magnetic field of a pulsar with $B=2\cdot 10^{13}$ G is about
$\delta\varphi^*\sim 2.1\cdot 10^{-4}$ rad $\approx  40$ arcsec according
to the quantum electrodynamics predictions and $\delta\varphi^*\sim
1.4\cdot 10^{-5}$ rad $\approx 2.8$ arcsec according to the BI
electrodynamics.

In the field of gamma-ray pulsar or magnetar with $B\sim 10^{16}$G the
maximum value of this angle increases:  $\delta\varphi^*\sim 3.5$ rad
$\approx 200^\circ $ according to the BI electrodynamics.

It is necessary to note that, the magnetic field of real neutron stars is
most certainly not dipolar (Feroci et al. 2001). In the case of a non
dipolar field the effective volume of nonlinear electrodynamics
interaction, will be smaller in comparison with the case of a dipolar
field. Thus, the gamma-ray bending will be less clearly pronounced.
However, the above obtained estimates made for the dipole field can be used
for model quantitative evaluations of nonlinear electrodynamics effects.

It is necessary to note that the obtained  values characterize deflection
angles at the source. At  the  point of   observation  possible  deflection
of  a  ray  will  be determined  by the attitude $\delta b/l_2,$ were
$\delta b$ is the variation of the impact parameter of a ray because of the
proper move of  the source  and  $l_2$ is the distance from the source to
observer. For typical values $l_2 = 10^{17}$ km (3 kps), $\delta b = 10^2$
km, $\delta b/l_2\sim 3.4\cdot 10^{-12}$  arcsec  that is beyond any
measurable limits now. The  only nonlinear  electrodynamics effect,  which
can  be  measured principally,  is the effect of gamma ray flux dispersion
by the  neutron star magnetic field. It follows from  equation  (20) that
flux  attenuation coefficient depends  on  the  impact parameter  $b$  as
well as on the ray bending  angle.  Thus, studying this effect we could
also obtain information on the nonlinear  electrodynamics bending of a ray
in  the  source.

The  effect  of gamma ray flux dispersion can be interpreted more
accurately, if the neutron star and the gamma ray source are  moving
relatively  each  other  with  periodical  time dependence of their mutual
location. As our estimates  show, the limits of the flux attenuation
coefficient value in this case  are  very  wide. Hence, in  the  case  of a
rotating  neutron  star,  if  gamma  rays are emitted  in   its vicinity,
we  will have periodic variations of the impact parameter which lead to the
periodic variations of the  flux attenuation  coefficient.  Such variations
will efficiently modulate the mean light curve, which can be detected  by
the outside observer. Under certain conditions the  same effect in the
orbital light curves could be observed in  the case of pulsars in binary
systems. However, the presence of  accreting  matter  can  lead  to  some
difficulties  in resolving the pure nonlinear electrodynamic effect from
the mean orbital light curves.

{\bf 
\subsection{The effect of birefringence}}

     The  main  qualitative difference  in the predictions  of different
nonlinear electrodynamics theories, which  can  be observed in  the
considered astrophysical conditions,  is the  absence  of  birefringence in
the  BI  theory.  According  to the HE theory vacuum birefringence effects
the polarization  of  the emitted photons  as  well  as  the radiative
opacity  of the near neutron  star  medium.  Such effects  leads  to  a
signature  in  the  emission  spectra.  However, for X-ray frequencies and
pulsar magnetic fields to be  quite pronounced this effect requires rather
high plasma density  ($10^{23}$  cm$^{-3}$)  (Meszaros $\&$ Ventura 1979),
which can take place in  some accretion   powered  pulsars.  In  the  case
of individual magnetic  neutron  stars (gamma-ray pulsars  and magnetars)
the magnetosphere  is quite transparent for hard emission.  The main
effect,  which  has influence on  the propagation  of emitted photons, is
absorption by pair production in a strong  magnetic field. This suppresses
the radiation  above about 1 MeV and effects the gamma ray beaming (Riffert
et al. 1989).

As for X- ray   frequencies   in  the  case  of  pulsars   with   poor
magnetosphere  we  suppose, that the only observable  difference between
predictions of BI and HE theories is that according to  the  first  theory
the nonlinear electrodynamic  bending angle  in  the dipole magnetic field
of a neutron  star  and consequently the flux attenuation coefficient  will
be  the same  for any polarization of electromagnetic waves. In  the HE
theory  $\eta_1\neq\eta_2$  as well as the nonlinear electromagnetic
bending   angle   and  consequently  the  flux   attenuation coefficient
depend on the electromagnetic wave polarization.

Thus,  this  theory  predicts that if the  gamma-ray  source moves
periodically relatively to the neutron star (i.e. in the case of rotating
neutron star or pulsar in binary system) we will obtain different  mean
light  curves  for   emitted electromagnetic  waves  with  perpendicular
mutual   linear polarization.

\section{Discussion}

Although the observation of the manifestations of the nonlinear
electrodynamics effects in astrophysical objects requires special
conditions, in principle, they can be observed. The main astrophysical
objects, where the nonlinear electrodynamic effects can be revealed more
clearly, are certain kinds of gamma-ray pulsars and magnetars. These
effects can be manifested as some peculiarities in the form of their hard
emission pulsation.

The typical luminosity of magnetars and certain kinds of rotation-powered
pulsars in hard emission is about $10^{34}-10^{36}$ erg s$^{-1}$
(Mereghetti 2000). For example, the sensitivity level of the Chandra
instruments for $10^4$ s exposure corresponds to the luminosity of galactic
objects $\sim 10^{30}-10^{31}$ erg s$^{-1}$ ( Garcia et al. 2001). Hence,
for the exposure time of about 1 s Chandra permits to obtain the mean light
curve of a source with the luminosity of $\sim 10^{32}-10^{33}$ erg
s$^{-1}$, this is about two order lower than the typical luminosity of such
magnetar-candidate sources as AXPs or SGRs.

The imaging mode of an instruments is not necessary for obtaining the mean
light curve, because for most of these objects (soft gamma-ray repeaters,
anomalous X- ray pulsars, etc) the pulsation period is known. However, the
detector outputs should be folded over the maximum possible time of
continuous exposure with that known period. For example, we can take $10^5
s$ as the exposure time of a given source, thus according to the announced
sensitivity for TeCd detector of the INTEGRAL IBIS instrument (Winkler
2001) we obtain the minimum detectable intensity of periodic processes (at
$8\sigma$ level), which is about 1 mCrab. This estimate permits to obtain
the mean curve of pulsation for a source like a SGR even in its quiescent
state. Since the number of known magnetar-like sources is not large, the
inspection of most of them during the mission will be quite justified.

Measurements of the bending angle in the source need observations of at
least the proper orbital motion of the pulsar, which is not accessible by
currently operating instruments.

The additional group of effects may be connected with such objects as
accretion-powered binary X-ray pulsars and, possibly other kinds of tight
binaries containing magnetic neutron stars. As it was mentioned above, the
objects with most suitable conditions for nonlinear electrodynamics effects
are binary systems with underfilled Roche lobe supergiants as an optical
companion.  The value of the  orbital period provides the necessary time
for continuous observations of these sources. Because  most of pulsar
systems with underfilled Roche lobe optical companion are characterized by
large orbital periods (4U1538-522, 3.73 d; 4U1907+097, 8.38 d;
1E1145.1-614, 5.648 d; Vela X-1, 8.965 d) (Bieldsten et al. 1997), rather
long exposures are necessary to obtain detailed mean orbital light curves.
The existense of certain non-typical forms can indicate the presence of
nonlinear electrodynamic effects in such objects.

\begin{acknowledgements}

We are grateful to Academician Georgiy T. Zatsepin for fruitful stimulating
discussions. Authors would also like to express their warm thanks to
Ekaterina D. Tolstaya for grammar corrections and Alexander A. Zubrilo,
Vitaly V. Bogomolov, Oleg V. Morozov for the help with preparing this
manuscript.

      Part of this work was supported by the Russian Foundation of Basic
Research.

\end{acknowledgements}

\appendix

\section{Eikonal equation }

Using Lagrangian (9), it is not difficult to obtain the expansion of
vectors $\bf D$ and $\bf H$ onto degrees $\xi $ with accuracy up to $\xi^2$
inclusive:  \begin{equation} {\bf D}={\bf E}+\xi \{2\eta_1 ({\bf E}^2-{\bf
B}^2){\bf E} +4\eta_2({\bf B\cdot E}){\bf B}\}+ \end{equation}
$$+\xi^2\{3\eta_3({\bf E}^2-{\bf B}^2)^2{\bf E}+ \eta_4({\bf B\cdot
E})^2{\bf E} +$$ $$+\eta_4({\bf E}^2-{\bf B}^2)({\bf B\cdot E}){\bf
B}\}\},$$ $${\bf H}={\bf B}+\xi \{2\eta_1({\bf E}^2-{\bf B}^2){\bf B}-
4\eta_2({\bf B\cdot E}){\bf E}\}+$$ $$+\xi^2\{3\eta_3({\bf E}^2-{\bf
B}^2)^2{\bf B} +\eta_4({\bf B\cdot E})^2{\bf B}-$$ $$-\eta_4({\bf E}^2-{\bf
B}^2)({\bf B\cdot E}){\bf E}\}.$$ Under the assumption of a "weak" plane
electromagnetic wave (8) with the use of an approximation linear in vectors
$\bf b$ and $\bf e$, we can obtain from relation (A1) and equations (1) a
uniform system of three linear algebraic equations relatively to three $\bf
e$ components:  \begin{equation} \Pi^{\alpha \beta }e_\beta =0,
\end{equation} where $$\Pi^{\alpha \beta }=A_0\{k^\alpha k^\beta
+({\omega^2\over c^2}-{\bf k}^2)\delta^{\alpha \beta }\}+$$
$$+A_1N_B^\alpha N_B^\beta +A_2{\omega^2\over c^2}B_0^\alpha B_0^\beta ,$$
and for more compact record we introduce the following notations:
$$A_0=1-2\eta_1\xi {\bf B}_0^2+3\eta_3\xi^2 {\bf B}_0^4,\ \ \
A_1=4\eta_1\xi -12\eta_3\xi^2 {\bf B}_0^2,$$ $$A_2=4\eta_2\xi
-\eta_4\xi^2{\bf B}_0^2,\ \ \ \ {\bf N}_B= [{\bf k\times  B}_0].$$ For the
existence of nontrivial solutions of the equations system (A2) it is
 necessary, that $det||\Pi^{\alpha \beta }||=0.$ The determinant of tensor
$\Pi^{\alpha \beta }$ can be calculated in the most simple way, if we use
the formulas of tensor algebra verified in the work of Denisova $\&$ Mehta
(1997).

The condition of equality to zero of the second order tensor determinant in
three-dimensional Euclidean space $\Pi^{\alpha \beta }$ can be rewritten
using these formulas in the form:
$$2\Pi_{(3)}-3\Pi_{(1)}\Pi_{(2)}+\Pi_{(1)}^{3}=0,$$ where
$$\Pi_{(N)}=\Pi_{\alpha_1 \beta_1}\delta_{ \beta_1\alpha_2} \Pi_{\alpha_2
\beta_2}\delta_{ \beta_2\alpha_3}\dots \Pi_{\alpha_N \beta_N}\delta_{
\beta_N\alpha_1}.$$ To compose the tensor's degrees $\Pi^{\alpha \beta },$
after the reduction by $\omega^2A_0/c^2$ we obtain the following dispersion
equation:  \begin{equation} {\omega^4 \over c^4}\Big\{A_2A_0{\bf B}_0^2
+A_0^2 \Big\}+ {\omega^2 \over c^2}\Big\{A_4[{\bf k\times B}_0]^2{\bf
B}_0^2- \end{equation} $$-A_2A_0{\bf B}_0^2{\bf k}^2 - 2A_0^2{\bf k}^2 +
A_1A_0[{\bf k\times B}_0]^2 -$$ $$-A_2A_0({\bf k\times B}_0)^2 \Big\}+
A_0^2{\bf k}^4 -A_1A_0{\bf k}^2[{\bf k\times B}_0]^2+$$ $$+A_2A_0 {\bf
k}^2({\bf k\cdot B}_0)^2- A_4[{\bf k\times B}_0]^2({\bf k\cdot B}_0)^2
=0,$$ where $A_4=A_1A_2-A_3^2.$

We will find the solution $\omega =\omega ({\bf k})$ of this equation as
the expansion onto degrees $\xi ,$ similar to the expansion of the
Lagrangian (9):  \begin{equation} \omega = ck[1+ \xi F+\xi^2U],
\end{equation} where $F$ and $ U$ are unknown functions.

Let us substitute now this expansion into equation (A3). Because the
Lagrangian (9) is written with accuracy up to terms proportional $\xi^2$
inclusive,  after calculations we can neglect the insignificant terms
$\xi^3.$ As a result we obtain a relation, which has the form of expansion
in degrees $\xi ,$ in view of its complication we will not write it here.

To make equal to zero the coefficients of this expansion in degrees $\xi ,$
we obtain the equation allowing to determine the unknown functions $F$ and
$ U.$ In the lowest approximation we have:  $$\{F k^2+2 \eta_1[{\bf k\times
B}_0]^2\} \{F k^2+ 2\eta_2 [{\bf k\times B}_0]^2\}=0.$$ To resolve this
equations relative to $F$ and substitute it by (A4), and obtain the
dispersion equations (10).

\section{ Calculation of the ray bending angle}

Differentiating the eikonal (15) with respect to $\alpha $
and making it equal to the constant $\beta ,$ we obtain
equation for a ray $\varphi =\varphi (r):$
\begin{equation}
\varphi =\beta + \alpha\int\limits^r{dr \over r^2
\sqrt{{{\cal E}_0^2\over c^2}\big[1+{4\eta_1\xi {\bf m}^2\over r^6}\big]
-{\alpha^2\over r^2}}}  .
\end{equation}
If we denote the source emission frequency
as $\omega_0$, then  the constant ${\cal E}_0=
\omega_0,$ and $\alpha =-\omega_0 b /c,$ where $b$ -
is the impact distance of the ray.

As a result expression (B1) becomes:
\begin{equation}
\varphi =\beta + b\int\limits^r{dr \over r^2
\sqrt{1+{4\eta_1\xi {\bf m}^2\over r^6}
-{b^2\over r^2}}}  .
\end{equation}
The integral in the right-hand part of this equality can be
expressed through elliptical functions. However, it is more
favorable for our purposes to find another approach. Let us
differentiate the equality (B2) with respect to $\varphi $ and
then to raise it into the second-order power:
\begin{equation}
\Big({dr\over d\varphi }\Big)^2={r^4\over b^2}\Big[
1+{4\eta_1\xi {\bf m}^2\over r^6}-{b^2\over r^2}\Big].
\end{equation}
To find the solution of this equation we will use the well
known Darwin method (1961).
First of all, we will introduce the subsidiary variable
$u=1/r.$ The equation (B3) becomes:
\begin{equation}
\Big({du\over d\varphi }\Big)^2={1\over b^2}\big[
1+4\eta_1\xi {\bf m}^2u^6-b^2u^2\big].
\end{equation}
We will find the solution of this equation in the form, as
\begin{equation} u=A+Q_1\sin \Psi(\varphi ),
\end{equation}
where $A$ and $Q_1$ are constants.

Substituting expression (B5) into equation (B4) and
restricting ourselves by the accuracy linear in the small
parameter $\eta_1\xi {\bf m}^2u^6,$ we obtain:
$$A=0,\ Q_1={1\over b}\big[1+{2\eta_1\xi {\bf m}^2\over b^6}\big],$$
and function $\Psi(\varphi )$ with acceptable accuracy
should satisfy equation
$${d\Psi(\varphi )\over d\varphi }=1-{2\eta_1\xi {\bf m}^2\over b^6}
\big[ 3-3\cos^2\varphi +\cos^4\varphi \big].$$
The solution of this equation is:
$$\Psi(\varphi )=\varphi -\varphi_0 -{\eta_1\xi {\bf m}^2\over 16 b^6}
\big[ 60\varphi -16\sin 2\varphi +\sin 4\varphi \big].$$
In the case, when the gamma-ray source is located at a
limited distance from a neutron star or even in its nearest
vicinity let us consider that this source is at the point
$r=l_1,\  \varphi =\pi .$
Then for a ray with impact distance $b$ the solution of
equation (B4) can be represented as:
$$u={1\over r}=Q_1\sin\big\{\varphi -\varphi_0
-{\eta_1\xi {\bf  m}^2\over 16 b^6}
\big[ 60\varphi -16\sin 2\varphi +\sin 4\varphi \big]\big\}.$$
Because this ray should pass through the point $r=l_1,\ \varphi =\pi ,$
the integration constant $\varphi_0 $ is more complicated in
comparison with expression (B6):
$$\varphi_0 =-{15\pi \eta_1\xi {\bf  m}^2\over 4 b^6}+
\arcsin({1\over Q_1l_1}).$$
The bending angle $\delta\varphi_1^*$ of a ray after its passing
through the neutron star magnetic field will be equal to:
$\delta\varphi_1^*=\varphi_1-\varphi_D,$ where $\varphi_1$ is the
angle of a ray inclination to the $X$ axis at the point
$r=l_1,\ \varphi =\pi ,$ and $\varphi_D$ is the angle of the
detected ray inclination to the $X$ axis.

Since in the considered case $l_2>>l_1,$ the $\varphi_D$
angle can be determined from the condition:
$u(\varphi_D)=1/l_2\approx 0.$
It follows from this, that
$$\varphi_D=-{15\pi \eta_1\xi {\bf  m}^2\over 4 b^6}+
\Big[1+{15 \eta_1\xi {\bf  m}^2\over 4 b^6}\Big]\arcsin({1\over Q_1l_1})+$$
$$+{\eta_1\xi {\bf  m}^2\over 16 b^6}\big\{\sin[4\arcsin({1\over Q_1l_1})]-
16\sin[2\arcsin({1\over Q_1l_1})]\big\}.$$
As it is well known, the tangens of $\varphi_1$
inclination angle at the point $r=l_1,\ \varphi =\pi ,$ is
equal to the derivative $dy/dx$ at this point:
$$tg (\varphi_1)={dy\over dx}(r=l_1, \varphi =\pi )=$$
$$={r'\sin\ \varphi +r\cos\ \varphi \over r'\cos\ \varphi -r\sin\ \varphi }
=-{1\over l_1u'}(\varphi =\pi ),$$
where the apostrophe denotes the derivative over $\varphi .$

In the first approximation in the small parameter $\xi {\bf m}^2/
b^6$ it can be obtained from this:
$$\varphi_1=\arctan\left[{bQ_1\over\sqrt{Q_1^2l_1^2-1}}\right].$$
As a result we obtain the formula (16)
for the bending angle $\delta\varphi_1^*$ of a ray.

\end{document}